\documentclass[sigconf]{acmart}
\usepackage{enumitem}
\usepackage{footnote}
\usepackage{amsmath}
\usepackage{array}

\usepackage{lscape}
\DeclareMathOperator*{\argmax}{argmax}
\def\BibTeX{{\rm B\kern-.05em{\sc i\kern-.025em b}\kern-.08emT\kern-.1667em\lower.7ex\hbox{E}\kern-.125emX}}

\copyrightyear{2019}
\acmYear{2019}
\setcopyright{acmlicensed}
\graphicspath{ {figures/} }
\begin{document}

%
\title{Neural Knowledge Extraction From Cloud Service Incidents}

%

\author{Manish Shetty}
\email{t-mashet@microsoft.com}
\affiliation{%
  \institution{Microsoft Research}
  \streetaddress{}
  \city{Bangalore}
  \state{India}
}

\author{Chetan Bansal}
\email{chetanb@microsoft.com}
\affiliation{%
  \institution{Microsoft Research}
  \streetaddress{}
  \city{Redmond}
  \state{USA}
}

\author{Sumit Kumar}
\email{sumiku@microsoft.com}
\affiliation{%
  \institution{Microsoft}
  \streetaddress{}
  \city{Redmond}
  \state{USA}
}

\author{Nikitha Rao}
\email{t-nirao@microsoft.com}
\affiliation{%
  \institution{Microsoft Research}
  \streetaddress{}
  \city{Bangalore}
  \state{India}
}

\author{Nachiappan Nagappan}
\email{nachin@microsoft.com}
\affiliation{%
  \institution{Microsoft Research}
  \streetaddress{}
  \city{Redmond}
  \state{USA}
}

\author{Thomas Zimmermann}
\email{tzimmer@microsoft.com}
\affiliation{%
  \institution{Microsoft Research}
  \streetaddress{}
  \city{Redmond}
  \state{USA}
}

%

\newcommand{\dummyfig}[1]{
  \centering
  \fbox{
    \begin{minipage}[c][0.1\textheight][c]{0.5\textwidth}
      \centering{#1}
    \end{minipage}
  }
}

\newcommand{\softner}{SoftNER}
\newcommand{\CompanyX}{Microsoft}
\newcommand{\todo}[1]{\textbf{\textcolor{red}{X #1 X}}}
\newcommand{\addRef}{\todo{REF HERE}}

%
\begin{abstract}
In the last decade, two paradigm shifts have reshaped the software industry - the move from boxed products to services and the widespread adoption of cloud computing. This has had a huge impact on the software development life cycle and the DevOps processes. Particularly, incident management has become critical for developing and operating large-scale services. Incidents are created to ensure timely communication of service issues and, also, their resolution. Prior work on incident management has been heavily focused on the challenges with incident triaging and de-duplication. 

In this work, we address the fundamental problem of structured knowledge extraction from service incidents. We have built \softner{}, a framework for unsupervised knowledge extraction from service incidents. We frame the knowledge extraction problem as a Named-Entity Recognition task for extracting factual information. \softner{} leverages structural patterns like key-value pairs and tables for bootstrapping the training data. Further, we build a novel multi-task learning based BiLSTM-CRF model which leverages not just the semantic context but also the data-types for named-entity extraction. We have deployed \softner{} at \CompanyX{}, a major cloud service provider and have evaluated it on more than 2 months of cloud incidents. We show that the unsupervised machine learning based approach has a high precision of 0.96. Our multi-task learning based deep learning model also outperforms the state of the art NER models. Lastly, using the knowledge extracted by \softner{} we are able to build significantly more accurate models for important downstream tasks like incident triaging.
\end{abstract}


%
\maketitle

\section{Introduction}
In the last decade, two major paradigm shifts have revolutionized the software industry. First is the move from boxed software products to \emph{services}. Large software organizations like Adobe, Autodesk\footnote{https://www.barrons.com/articles/autodesks-bet-on-the-cloud-will-generate-big-returns-for-shareholders-1443248123} and Microsoft\footnote{https://www.pcworld.com/article/2038194/microsoft-says-its-boxed-software-probably-will-be-gone-within-a-decade.html} which pre-date the internet revolution have been aggressively trying to move from selling boxed products to subscription based services. This has primarily been driven by the benefits of subscription based services. In terms of technical advantages, these services can always be kept up-to-date, they do not need to wait for the next shipping cycle to install security patches, and, lastly, the telemetry from these services provide invaluable insights about the usage of the products. On the monetary side, these subscription based services provide recurring revenue, instead of a one time sale. Also, it allows companies to reach their customers directly, bypassing the vendors and retailers. The second major shift has been the widespread adoption of \emph{public clouds}. More and more software companies are moving from on-premises data centers to public clouds like Amazon AWS, Google Cloud, Microsoft Azure, etc. Gartner has forecasted\footnote{https://www.gartner.com/en/newsroom/press-releases/2019-11-13-gartner-forecasts-worldwide-public-cloud-revenue-to-grow-17-percent-in-2020} the public cloud market to grow to about \$$266$ billion revenue in $2020$, out of which about 43\% revenue will be from the Software as a Service (SaaS) segment. The cloud revolution has enabled companies like Netflix, Uber, etc. to build internet scale products without having to stand up their own infrastructure.

While the paradigm shifts have revolutionized the software industry, they have also had a transformational effect on the way software is developed, deployed and maintained. For instance, software engineers no longer develop monolithic software. They build services which have dependencies on several 1st part and 3rd party services and APIs. Typically, any web application will leverage cloud services for basic building blocks like storage (relational and blob), compute, authentication. These complex dependencies introduce a bottleneck where a single failure can have a cascading effect. In 2017, a small typo led to a major outage in the AWS S3 storage service, which ended up costing over \$$150$ million to customers like Slack, Medium, etc\footnote{https://www.wsj.com/articles/amazon-finds-the-cause-of-its-aws-outage-a-typo-1488490506}. Similarly, Google Cloud, had a config error affect Google cloud for 4+ hours \footnote{https://www.zdnet.com/article/google-details-catastrophic-cloud-outage-events-promises-to-do-better-next-time/}. Microsoft had a glitch in their Active Directory last October\footnote{https://www.theregister.co.uk/2019/10/18/microsoft\_azure\_mfa/}, which locked out customers from accessing their Office 365 and Azure accounts. These outages are inevitable and can be caused by various factors such as code bugs, mis-configurations \cite{mehta2020rex} or even environmental factors.

To keep up with these changes, DevOps processes and platforms have also evolved over time \cite{lipredicting, dang2019aiops, kumar2019building}. Most software companies these days have incident management and on-call duty as a part of the DevOps workflow. Feature teams have on-call rotations where engineers building the features take on incident management responsibilities for their products. Depending on the severity of the incident and the service level objective (SLO), they would need to be available 24x7 and respond to incidents in near-real time. The key idea behind incident management is to reduce impact on customers by mitigating the issue as soon as possible. We discuss the incident life-cycle and some of the associated challenges in detail in Section~2. Prior work on incident management has largely focused on two challenges: incident triaging \cite{ContinuousTriageASE2019, EmpiricalIcMICSE2019} and diagnosis \cite{nair2015learning, bansal2019decaf, luo2014correlating}. Chen et al. \cite{ContinuousTriageASE2019} did an empirical study where they found that upto 60\% of the incidents can be mis-triaged. They proposed DeepCT, a deep learning approach for automated incident triaging using incident data (title, summary, comments) and the environmental factors. 

In this work, we address the complimentary problem of \textbf{extracting structured knowledge from service incidents}. Based on our experience from operating web-scale cloud services at \CompanyX{}, incidents created by internal and external customers contain huge amount of unstructured information. Ideally, any knowledge extraction framework should have the following qualities:
\begin{enumerate}
    \item It should be \textbf{unsupervised} because it's laborious and expensive to annotate a large amount of training data. This is important since every service and team would have it's own vocabulary. For instance, the information or entities contained in incidents for the storage service is very different from the incidents for the compute service.
    \item It should be \textbf{domain agnostic} so that it can scale to a high volume and a large number of entity types. In the web domain, there are a small set of key entities such as people, places, organizations. However, in the incidents, we don't know these entities apriori.
    \item It should be \textbf{extensible}, we should be able to adapt the bootstrapping techniques to incidents from other services or even other data-sets such as bug reports. This is critical because each service could have it's unique vocabulary and terminology.
\end{enumerate}

We have designed \textbf{Soft}ware artifact \textbf{N}amed-\textbf{E}ntity \textbf{R}ecognition (\softner{}), a framework for unsupervised knowledge extraction from service incidents, which has these three qualities: unsupervied, domain agnostic, and extensible. We frame the knowledge extraction problem as a \textit{Named-entity recognition} task which has been well explored in the Information Retrieval domain \cite{nadeau2007survey, lample2016neural}. It leverages structural pattern extractors for bootstrapping the training data for knowledge extraction. Further, \softner{} incorporates a novel multi-task BiLSTM deep learning model with attention mechanism. We have evaluated and deployed SoftNER at \CompanyX{}, a major cloud service provider. We show that our unsupervised entity extraction has high precision. Our multi-task deep learning model also outperforms existing state of the art models on the NER task. Lastly, using the extracted knowledge, we are able to build more accurate models for key downstream tasks like incident triaging. In this work, we make the following main contributions:
\begin{enumerate}
    \item We propose \softner{}, the first approach for completely unsupervised knowledge extraction from service incidents.
    \item We build a novel multi-task learning based deep learning model which leverages not just the semantic features but also the data-types. It outperforms the existing state of the art NER models. 
    \item We do an extensive evaluation of \softner{} on over 2 months of cloud service incidents from \CompanyX{}\footnote{We cannot disclose the number of incidents due to Microsoft policy.}.
    \item Lastly, we have deployed \softner{} in production at \CompanyX{} where it has been used for knowledge extraction from over 400 incidents.
\end{enumerate}

The rest of the paper is organized as follows: In Section 2, we discuss insights from the incident management processes and the challenges at \CompanyX{}. Section 3 and 4 provide an overview, and details of our approach for knowledge extraction from incidents with \softner{}, respectively. In Section 5, we discuss the experimental evaluation of our approach. Section 6 describes the various applications and Section 7 discusses the related work. We conclude the paper in Section 8.
\section{Incident life-cycle}
\label{incident-life-cycle}

\begin{figure}[H]
\vspace{-6pt}
\includegraphics[width=\linewidth]{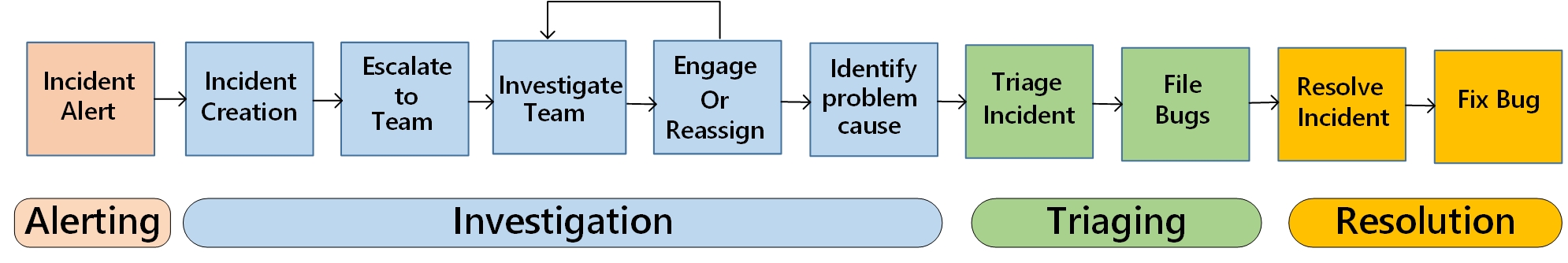}
\caption{Incident Life-Cycle}
\label{fig:incident-lifecycle}
\vspace{-6pt}
\end{figure}

In \CompanyX{}, an incident is defined as an unplanned interruption or degradation of a product or service that is causing customer impact. For example, a slow connection, a timeout, a crash etc. could constitute an incident. The core part of our paper is to understand the incident management process which defines the various steps an incident goes through - from creation to closing. Bugs and incidents are very different as in our case incidents may or may not lead to bugs. Furthermore, incidents often require the involvement of developers on-call who have been designated to respond to incidents. Figure \ref{fig:incident-lifecycle} presents a high level view of the incident management process. Most online services have their own specific incident management protocol. This figure is a generic process that should apply outside of \CompanyX{} as well.

The incident management process is broadly classified into 4 phases. In the first phase which is the \emph{alerting} phase, typically an alert is fired when the service monitoring metrics fall below a pre-definitely acceptance level in terms of performance (for example slow response), slow transfer rate, system hang or crash, etc. This leads to phase 2 which is the \emph{investigation} phase. In the investigation phase, firstly an incident is created in the incident database. This is then escalated to a "related" team. The identification of the first "related" team is automatic, based on heuristics or component ownership. The team investigates the incidents and engages with relevant stakeholders or re-routes it to the appropriate team to repeat the steps. 
    
Our work applies directly in the engagement and problem identification phase dealing with unsupervised knowledge extraction from service incidents to aid in the problem resolution. Then the appropriate team identifies the problem cause and moves over to the next phase, which is called \emph{triaging}. In this phase, the incident is triaged according to priority and the appropriate bugs for fixing the incident are filed for the engineering teams to fix. In the final phase of \emph{resolution}, the incident is resolved and the bug is fixed in the system. There are other activities including root cause analysis which happen outside of this timeline in parallel in order to ensure that incidents do not repeat in the future.
\section{Overview}

\begin{figure*}
\includegraphics[width=1\textwidth]{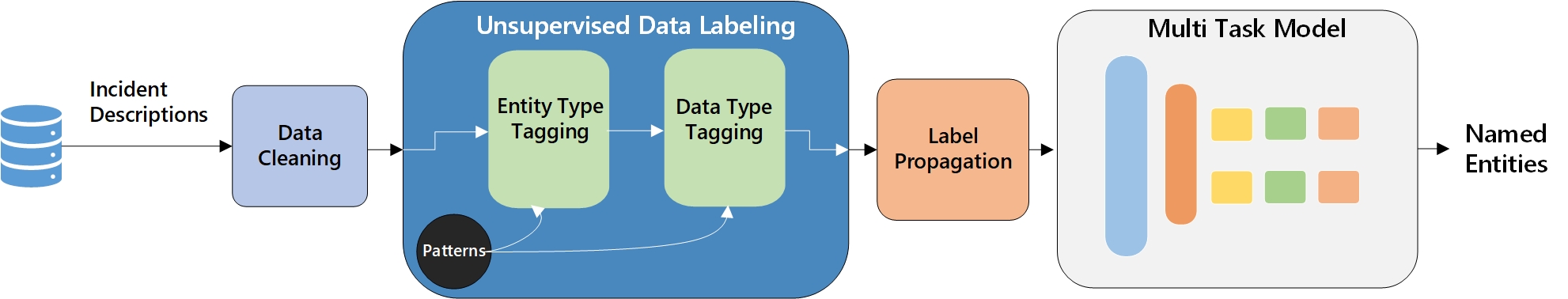}
\caption{Machine learning pipeline}
\label{fig:overview}
\end{figure*}

Incident management is key to running large scale services in an efficient and performant way. However, there is a lot of scope for optimization which can increase customer satisfaction, reduce on-call fatigue and provide revenue savings. Existing work on incident management has focused on specific problems such as incident triaging. The state of the art incident triaging methods \cite{ContinuousTriageASE2019} use novel deep learning methods which takes the unstructured incident description, discussions and title, and predict the team to which the incident should be triaged. In this work, we focus on the fundamental problem of structured knowledge extraction from these unstructured incidents. With the structured information, not only can we build machine learning models for tasks like triaging, but we can also save a lot of time and effort for on-call engineers by automating the manual processes such as running health checks on resources.

To solve the challenges with incident management, we have designed the \softner{} framework. It is the first automated approach for structured knowledge extraction from service incidents. We frame the knowledge extraction problem as a \textit{Named-Entity Recognition} (NER) task which has been well explored in the Information Retrieval (IR) domain \cite{nadeau2007survey, lample2016neural}. Named-Entity Recognition is defined as the task of parsing unstructured text to not only detect entities but also classify them into specific categories. An entity can be any chunk of text which belongs to a given type or category. For example, in the IR domain, some of the most commonly used entity types are people, locations and organizations. As an example, here is the input and output of a NER task for a news headline:

\smallskip
\textit{\textbf{Input}}: Over 320 million people have visited the Disneyland in Paris since it opened in 1992.

\smallskip
\textit{\textbf{Output}}: Over \textbf{[\textsubscript{COUNT} 320 million]} people have visited the \textbf{[\textsubscript{ORG} Disneyland]} in \textbf{[\textsubscript{LOC} Paris]} since it opened in \textbf{[\textsubscript{YEAR} 1992]}.

\smallskip

Framing the knowledge extraction problem as a NER task enables us to not only extract factual information from the incidents but also classify them as specific entities. For instance, if we just extract a \textit{GUID} from the text, it provides limited context. However, identifying that \textit{GUID} as a \textit{Subscription Id} is much more useful. Figure \ref{fig:overview} provides an overview of the various components of the \softner{} framework. In this framework we combine techniques from the information retrieval and the deep learning domains for solving the problem of knowledge extraction. One key limitation of any machine learning based pipeline is the requirement of huge amounts of labeled data which can be cost prohibitive to manually generate. In service incidents, the lack of existing training data prevents us from using any supervised or semi-supervised techniques.

\softner{} uses pattern extractors which leverage the \emph{key-value} and \emph{tabular} structural patterns in the incident descriptions to bootstrap the training data. We then use label propagation to generalize the training data beyond these patterns. Then we design a novel Multi-Task deep learning model which is able to extract named-entities from the unstructured incident descriptions with very high accuracy. The model not only leverages the semantic features but also the data type of the individual tokens (such as GUID, URI, Boolean, Numerical, IP Address etc.). Below we list some of the terms which we will be using throughout the rest of the paper. These terms identify different aspects of the named-entities.

\begin{itemize}
\item \textbf{Entity Name:} N-gram indicating the name of the entity. In the current implementation, N can range from 1 to 3. We also enforce the entity names to be alphabetical to filter out noisy data.
\item \textbf{Entity Value:} The value of the named-entity for a given instance. 
\item \textbf{Data Type:} The data type of the values for the named-entity.
\end{itemize} 

\textbf{Example}: Let's consider a real incident reported by a customer of the Cloud Networking service operated by \CompanyX{}. The incident was caused due to an error in deleting a Virtual Network resource. The key information required to triage and mitigate this incident is actually the `\textit{Problem Type}' and the `\textit{Virtual Network Id}'. This information is already present in unstructured form within the incident description, the challenge is to extract it automatically in a structured format. Using \softner{}, we can not only provide key information about the incidents to the on-call engineers, but, also automate some of the manual tasks such as running health checks. For instance, in the previous example, we can automatically look up the current status and logs before the on-call engineer engages. Lastly, the extracted knowledge can also be used to build more accurate machine learning models for predictive tasks like triaging, root causing, etc.
\section{Our approach}
Next, we describe our approach in implementing the \softner{} framework in detail. As shown in Figure \ref{fig:overview}, we start with the data cleaning process, followed by unsupervised data labeling. Then we describe the label propagation process and the architecture of the deep learning model.

\subsection{Data Cleaning}
Service incident descriptions and summaries are created by various sources such as external customers, feature engineers and even automated monitoring systems. The information could be in various forms, like textual statements, conversations, stack traces, shell scripts, images,  etc., all of which make the information unstructured and hard to interpret. Yet, these descriptions are a goldmine of information, in the form of identifiable entities, amidst other less useful information. Here, we describe the different approaches taken to clean the data before extracting information. First, we prune tables in the incident description that have more than 2 columns and get rid of HTML tags using regexes and HTML parsers. In this process, we also segment the information into sentences using newline characters. Next, we process individual sentences by cleaning up extra spaces and tokenize them into words. Our tokenization technique is able to handle camel-case tokens and URLs as well.

\begin{table*}[!ht]
\small
\caption{Examples of entities extracted by \softner{}}
\vspace{-6pt}
\label{entity-examples}
\begin{tabular}{lll}
\toprule
\textbf{Entity Name} & \textbf{Data Type} & \textbf{Example} \\
\midrule
Problem Type & Alphabetical & VNet Failure \\[-1pt]
Exception Message & Alphabetical & The vpn gateway deployment operation failed due to an intermittent error \\[-1pt]
Failed Operation Name & Alphabetical & Create and Mount Volume \\[-1pt]
Resource Id & URI & /resource/2aa3abc0-7986-1abc-a98b-443fd7245e6f/resourcegroups/cs-net/providers/network/frontdoor/ \\[-1pt]
Tenant Id & GUID & 4536dcd6-e2e1-3465-a22b-d25f62456233 \\[-1pt]
Vnet Id & GUID & 45ea1234-123b-7969-adaf-e0255045569e \\[-1pt]
Link With Details & URI & https://supportcenter.cloudx.com/caseoverview?srid=112\\[-1pt]
Device Name & Other & sab01-98cba-1d \\[-1pt]
Source IP & IP Address & 198.168.0.1 \\[-1pt]
Status Code & Numeric & 500 \\[-1pt]
Location & AlphaNumeric & eastus2\\[-1pt]
\bottomrule
\end{tabular}
\end{table*}

\subsection{Unsupervised Data Labelling}
For a lot of tasks in data mining like sentiment classification or even NER, supervised or semi-supervised methods are generally used. However, we don't have any pre-existing labelled data set which can  be used for a supervised NER task. It would also be very expensive to generate labelled data since the entity types vary across different services. So, we have built \softner{} as a completely unsupervised framework. Please refer to Table \ref{entity-examples} for examples of entities extracted using the unsupervised approach. Here are the steps for automatically generating the labelled corpus for named-entity extraction:

\textbf{Step 1 (Entity tagging)}: Since we don't have a list of entity types apriori, we first bootstrap the framework with a candidate set of entity name and value pairs. For this, we have built pattern extractors using the structural patterns commonly used in the incident descriptions: 
\begin{itemize}
    \item \textbf{Key-Value pairs} - This pattern is commonly used in the incident descriptions to specify various entities where the Entity Type and Value are joined by a separator such as ':'. For instance, "\textit{Status code: 401}" or "\textit{Problem type: VM not found}". Here, we split the sentence on the separator and extract the first half as the \textit{Entity Type} and the second half as the \textit{Entity Value}.
    \item \textbf{Tables} - Tables also occur quite frequently in the incident descriptions, especially, the ones which are created by bots or monitoring services. We extract the text in the header tags '\textit{<th>}' as the \textit{Entity Types} and the values in the corresponding rows as the \textit{Entity Values}.
\end{itemize}
 
\textbf{Step 2}: Now, we have a candidate set of entity names and values. However, the candidate set is noisy since we have extracted all the text which satisfies these patterns. In the NER task, entity name corresponds to the category names (for instance, people, location, etc.). So, we filter out any candidates where the entity name contains symbols or numbers. Also, for any NER framework, it's important to have a robust set of named-entities. So, we extract n-grams (n: 1 to 3) from the entity names of the candidates and take the top 100 most frequently occurring n-grams. In this process less frequent, thus noisy candidate entity types, such as \textit{"token acquisition started"}, are pruned. Also with this n-gram analysis, a candidate entity such as [\textit{"My Subscription Id is", "6572"}] would be transformed to [\textit{"Subscription Id", "6572"}] since \textit{"Subscription Id"} is a commonly occurring bi-gram in the candidate set. 

\textbf{Step 3 (Data-type tagging)}: For the refined candidate set, we next infer the data type of the entity values using in-built functions of Python such as "isnumeric" along with regexes. We leverage multi-task learning in \softner{}, where we jointly train the model to predict both the entity type and the data type. These tasks are complementary and help improve the accuracy for the individual prediction tasks. Based on discussions with the service engineers, we have defined the following data types:
\begin{itemize}
    \item \textbf{Basic Types}: Numeric, Boolean, Alphabetical, Alphanumeric, Non-Alphanumeric
    \item \textbf{Complex Types}: GUID, URI, IP Address
    \item \textbf{Other}
\end{itemize}
To infer the data type, we compute it for each instance of a named entity. Then, conflicts are resolved by taking the most frequent type. For instance, if "VM IP" entity is most commonly specified as an IP Address but sometimes is specified as a boolean, due to noise or dummy values, we infer it's data type as an IP Address.

\subsection{Label Propagation}
With the unsupervised tagging, we have bootstrapped the training data using the pattern extraction. While this allows us to generate a seed dataset, the recall would suffer since the entities could occur inline within the incident descriptions without the key-value or tabular patterns. In the absence of ground truth or labeled data, it's a non-trivial problem to solve. In \softner{} we use the unsupervised techniques to label the incident descriptions which are then used to train a deep learning based model. So, to avoid over-fitting the model on the specific patterns, we would want to generalize or diversify the labels.

We use the process of label propagation to solve this challenge. We use the entity values extracted in the bootstrapping process and propagate their types to the entire corpus. For instance, if the IP Address "127.0.0.1" was extracted as a "Source IP" entity, we would tag all un-tagged occurrences of "127.0.0.1" in the corpus as "Source IP". As we can imagine, there are certain edge cases that need to be handled. For instance, we cannot use this technique for entities with Boolean data type. It would also not work for all multi token entities, particularly, the ones which are descriptive. Lastly, it's possible that different occurrences of a particular value were tagged as different entities during bootstrapping. For instance, "127.0.0.1" can be "Source IP" in one incident while "Destination IP" in another incident. We resolve conflicts during label propagation based on popularity, i.e., the value is tagged with the entity type which occurs more frequently across the corpus.

\newcommand{\addREf}{\todo{REF HERE}}

\subsection{\softner{} Deep Learning Model}
\label{sec:multi-task-model}
The previous sections explain the phases of the \softner{} framework, as shown in Figure \ref{fig:overview}, that automate the significant task of creating labeled data for deep learning models which can further generalize knowledge extraction. Here we propose a novel Multi Task deep learning model that solves two entity recognition tasks simultaneously, i.e., Entity Type recognition and Data Type recognition.

The model uses an architecture, as described in Figure \ref{model-arch}, that shares some common parameters and layers for both tasks, but also has task specific layers. Incident descriptions are converted to word level vectors using a pre-trained Glove Embedding layer. This sequence of vectors is interpreted, both forwards and in reverse, by a Bi-directional LSTM layer. We then have distinct layers for the two tasks. The time-distributed layer transposes the BiLSTM hidden vectors to the shape of the output labels and the attention mechanism helps the model bias it's learning towards important sections of the sentences. Finally, the CRF layer produces a valid sequence of output labels. We perform back propagation using a combination of loss functions during training and evaluate individual tag precision, recall and F1 metrics. In the following sub sections we describe the important layers and approaches used in our multi task model.

\subsubsection{\textbf{Word Embeddings}}
\label{sec:glove-word-emb}
Language models in the semantic vector space, require real valued vectors as word representations. The importance of these vectors in improving performance for NLP tasks has been widely studied \cite{collobert2011natural}. These vectors act as characteristic features in applications of language models like question answering, document classification and named entity recognition.
GloVe \cite{pennington2014glove}, introduced a model that captures linear substructure relations in a global corpus of words, revealing regularities in syntax as well as semantics. The GloVe model trained on 5 different corpora, covers a vast range of topics and tokens. GloVe vectors, demonstrated on tasks such as word analogy and named entity recognition in \cite{pennington2014glove}, outperforms various other word representations. To use the 100 dimension version of GloVe, we create an embedding layer with the pre-trained GloVe weights in all our models.

\subsubsection{\textbf{Bi-directional LSTM}}
Recurrent Neural Networks (RNN) have been the basis for numerous language modelling tasks in the past.\cite{mikolov2010recurrent}. An RNN maintains historic information extracted from sequence or series like data. This feature enables RNN based models to make predictions at a certain time step, conditional to the viewed history. They take a sequence of vectors $\boldsymbol{(x_1, x_2, .., x_n)}$ as input and return a sequence of vectors $\boldsymbol{(h_1, h_2, .., h_3)}$ that encodes information at every time step.
Although it can be hypothesized that RNNs are capable of encoding and learning dependencies that are spread over long time steps, evidence shows that they fail to do so. RNNs tend to be biased towards more recent updates in long sequence situations.

Long Short-term Memory (LSTM) networks \cite{hochreiter1997long} were designed to overcome the problems associated with vanilla RNNs. Their architecture allows them to capture long range dependencies using several gates. These gates control the portion of the input to give to the memory cell, and the portion from the previous hidden state to forget. We model the LSTM layer using the following equations:
\begin{equation}
    f_{t}=\sigma(W_{f}\cdot[h_{t-1},x_{t}] + b_{f})
\end{equation}
\begin{equation}
    i_{t}=\sigma(W_{i}\cdot[h_{t-1},x_{t}] + b_{i})
\end{equation}
\begin{equation}
    {\tilde {c}}_{t}=\tanh(W_{c}\cdot[h_{t-1},x_{t}] + b_{c})
\end{equation}
\begin{equation}
    c_{t}= f_{t}\circ c_{t-1}+ i_{t}\circ {\tilde {c}}_{t}
\end{equation}
\begin{equation}
    o_{t}=\sigma(W_{o}\cdot[h_{t-1},x_{t}] + b_{o})
\end{equation}
\begin{equation}
    h_{t}=o_{t}\circ \tanh(c_{t})
\end{equation}

In the above equations $\sigma$ is the element wise sigmoid function and the $\circ$ represents hadamard product (element-wise). $f_t, i_t$ and $o_t$ are forget, input, and output gate vectors respectively, and, $c_t$ is the cell state vector. Using the above equations, given a sentence as a sequence of real valued vectors $\boldsymbol{(x_1, x_2, .., x_n)}$, it computes $\overrightarrow{h_{t}}$ that represents the leftward context of the word at the current time step $t$. By intuition, a word at the current time step $t$, receives context from other words that occur on either sides. A representation of this can be achieved with a second LSTM that interprets the same sequence in reverse, returning $\overleftarrow{h_{t}}$ at each time step. This combination of forward and backward LSTM is referred to as Bi-Directional LSTM (BiLSTM) \cite{graves2005framewise}. The final representation of the word is produced by concatenating the left and right context, $h_{t}=[\overrightarrow{h_{t}};\overleftarrow{h_{t}}]$.

\subsubsection{\textbf{Neural Attention Mechanism}}
In recent years attention mechanism has become increasing popular in various NLP applications like neural machine translation \cite{bahdanau2014neural}, sentiment classification \cite{chen2017recurrent} and parsing \cite{li2016discourse}. Novel architectures like transformers \cite{vaswani2017attention} and BERT \cite{devlin2018bert} have proven the effectiveness of such a mechanism for various downstream tasks. In addition to the improvements BiLSTMs bring to the original RNN approach, attention mechanism addresses long input sequences by retaining and utilising all hidden states generated. We implement attention at the word level as a neural layer, with a weight parameter $W_{a}$. It takes as input the the hidden states from the BiLSTM, transposed to output dimensions using a time distributed dense layer. Let $h = [h_1, h_2, .. h_{T}]$ be the input to the attention layer. The attention weights and final representation $h^*$ of the sentence is formed as follows:
\begin{equation}
    scores = W_a^Th
\end{equation}
\begin{equation}\label{attention-score}
    \alpha = softmax(scores)      
\end{equation}
\begin{equation}
    r = h\alpha^T    
\end{equation}
\begin{equation}\label{attention-repr}
    h^* = \tanh(r)    
\end{equation}

\begin{figure}
\includegraphics[width=\linewidth]{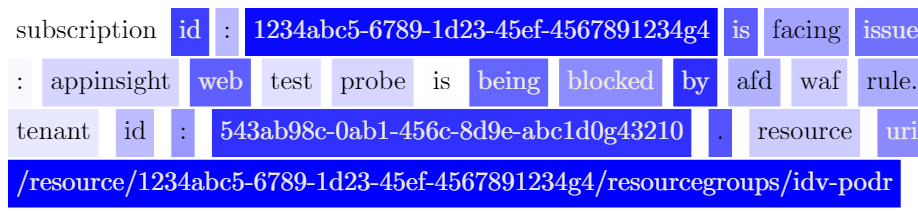}
\vspace{-12pt}
\caption{Attention visualization on a sample input}
\label{attention-viz}
\vspace{-6pt}
\end{figure}

In equations \ref{attention-score} \& \ref{attention-repr} the $softmax$ and $tanh$ functions are applied element-wise on the input vectors. We then concatenate $h$ and $h^*$ and pass it to the next layer. We visualize the attention vector $\alpha$ for a test sentence in Figure \ref{attention-viz}, where we observe that the attention layer learns to give more emphasis to tokens that have a higher likelihood of being entities. In Figure \ref{attention-viz}, the darkness in the shade of blue is proportional to the degree of attention. In case of long sequences, this weighted attention to certain sections of the sequence, that are more likely to contain entities, helps improve the model's sensitivity\footnote{Sensitivity, also known as recall, is the proportion of actual positives that are correctly identified}.

\subsubsection{\textbf{Conditional Random Fields}}
A simple approach to sequence tagging with LSTMs is to use the hidden state representations ($h_t$) as word features to make independent tagging decisions at the word level. But that leaves behind inherent dependencies across output labels in tasks like Named Entity Recognition. Our NER task also has this characteristic since the initial \softner{} heuristics enforce structural constraints, for example, separators between key-value and html table tags. In learning these dependencies and generalizing them to sentences without these constraints, we model tagging decisions jointly using conditional random fields 
\cite{lafferty2001conditional}.

For an input sequence $\boldsymbol{X = (x_1, x_2, .., x_n)}$, let $\boldsymbol{y = (y_1, y_2, .., y_n)}$ a potential output sequence, where n is the no of words in the sentence. Let $P$, the output of the BiLSTM network passed through the dense and attention layers, be the matrix of probability scores of shape $n \times k$, where k is the number of distinct tags. That is $P_{i,j}$ is a score that the $i^{th}$ word corresponds to the $j^{th}$ tag. We define CRF as a layer in the model, whose working is as follows. First a score is computed for $y$.

\begin{equation}
    s(\boldsymbol{X,y}) = \sum_{i=0}^{n} A_{y_i,y_{i+1}} + \sum_{i=0}^{n} P_{i,y_{i}} 
\end{equation}

where A represents the matrix of transition scores. That is $A_{i,j}$ is the score for the transition from $tag_i$ to $tag_j$. Then the score is converted to a probability for the sequence $y$ to be the right output using a softmax over $\boldsymbol{Y}$(all possible output sequences). 

\begin{equation}
    p(\boldsymbol{y|X}) = 
    \dfrac{e^{s(\boldsymbol{X,y})}}{\sum_{y' \in Y} e^{s(\boldsymbol{X,y'})}  }
\end{equation}

The model learns by maximizing the log-probability of the correct $y$. While extracting the tags for the input, we predict the output sequence with the highest score.

\begin{equation}
    \boldsymbol{y^*} = \argmax_{y' \in Y} p(y'|X)
\end{equation}

From the above implementation it is clear as to how the CRF and attention layers push the model towards learning a valid sequence of tags, unlike in the case of independent tagging as discussed.

\begin{figure}
\includegraphics[width=\linewidth]{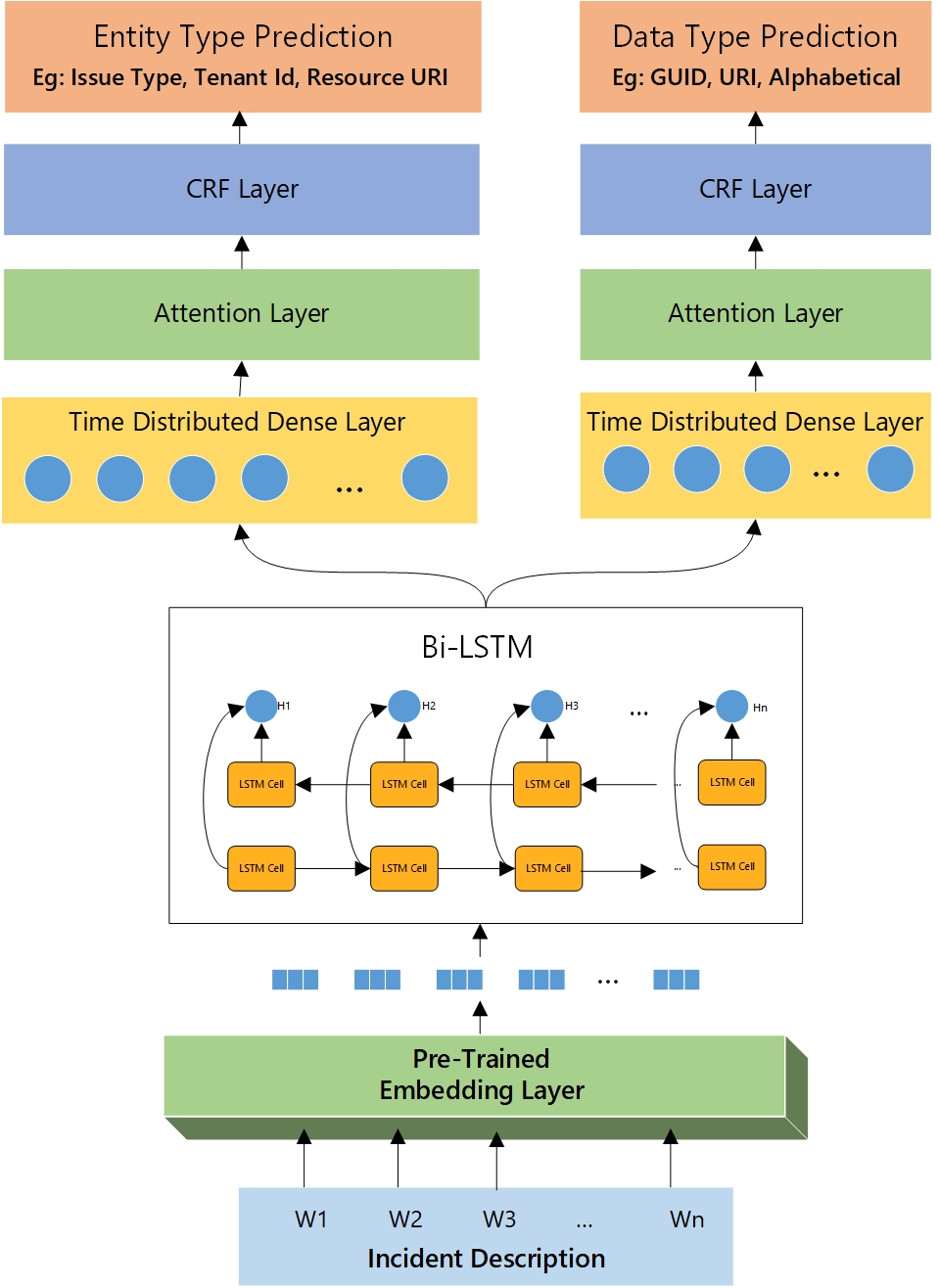}
\caption{Multi-task model architecture}
\label{model-arch}
\end{figure}

\subsubsection{\textbf{Multi-Task Learning}}

Caruana et al. \cite{caruana1997multitask} defines Multi-Task Learning (MTL) as an approach to improve generalization in models by using the underlying common information contained among related tasks. Some well known applications for MTL are multi-class and multi-label classification problems. In the context of classification or sequence labelling, MTL improves the performance of individual tasks by learning them jointly.

In \softner{}, named-entity recognition is the primary task. In this task, models mainly learn from context words that support occurrences of entities. But we also observe that incorporating a complimentary task of predicting the data-type of a token reinforces intuitive constraints, indirectly to model training. For example in an input like "\textit{The SourceIPAddress is 127.0.0.1}", the token \textit{127.0.0.1} is identified more accurately by our model, as the entity type "\textbf{Source Ip Address}", because it is also identified as the data-type "\textbf{Ip Address}", in parallel. This supplements the intuition that all \textit{Source Ip Addresses} are \textit{Ip adresses}, thus, improving model performance. Therefore, we choose data type prediction as the auxiliary task for \softner{}'s deep learning model.

There are multiple architectures that allow multi-task learning, like Multi-head architecture, Cross-snitch Networks\cite{misra2016cross} and Sluice Networks \cite{ruder2017learning}. We use the Multi-head architecture, where the lower level features generated by the BiLSTM layers are shared, whereas the other layers are task specific. The combined architecture is depicted in Figure \ref{model-arch}. As stated above, we define the entity type prediction as the main task and that of data type prediction as the auxiliary task. The losses are initially calculated individually for both tasks, $l_1$ and $l_2$, and then combined into $loss_{c}$ using a weighted sum. The parameter $loss\_weights = (\alpha,\beta)$ is used to control the importance between main and auxiliary task as follows:

\begin{equation}
    loss_{c} = \alpha \times l_1 + \beta \times l_2
\end{equation}

During training, the model aims to minimize the $loss_{c}$ but the individual losses are back-propagated to only those layers that produced the output. With such an approach, the lower level common layers are trained by both tasks, whereas the task specific layers are trained by individual losses.

\subsection{Implementation}

We implement \softner{} and all the machine learning models using Python 3.7.5, with Keras-2.2.4 and the tensorflow-1.15.0 backend. The hyper-parameters for the deep learning models are set as follows: word embedding size is set to 100, hidden LSTM layer size is set to 200 cells, and, maximum length of a sequence is limited to 300. These hyper-parameters were re-used among all models. The embedding layer uses pre-trained weights from glove.6B.100d, downloaded from the official stanford-nlp website\footnote{http://nlp.stanford.edu/data/glove.6B.zip}. Our models are trained on an Ubuntu 16.04 LTS machine, with 24-core Intel Xeon E5-2690 v3 CPU (2.60GHz), 112 GB memory and 64-bit operating system. The machine also has a Nvidia Tesla P100 GPU with 16 GB RAM.

We have also deployed \softner{} as a REST API developed using the Python Flask web app framework. The REST API offers a POST endpoint which takes the incident description as input and returns the extracted entities in JSON format. We have deployed it on the cloud platform provided by \CompanyX{} which allows us to automatically scale the service based on the variation in request volume. This enables the service to be cost efficient since majority of the incidents are created during the day. We have also enabled application monitoring which alerts us in case the availability or the latency regresses.
\section{Evaluation}
\softner{} solves the problem of entity recognition and extraction from unstructured text descriptions of incidents. But to evaluate the \softner{} framework in its entirety, we propose a 3 phase evaluation, to assess \softner{}'s applicability to incident management. We thus, answer the following questions using the evaluation:
\begin{itemize}
    \item \textbf{Entity Types}: How does \softner{}'s unsupervised approach perform in recognizing distinct entity types?
    \item \textbf{SoftNER Model}: How does the \softner{}'s Multi-Task model compare to state-of-the-art deep learning approaches for the NER task?
    \item \textbf{Auto-Triage}: Does \softner{} help improve the downstream task of automated incident triaging?
\end{itemize}

\subsection{Study Data}
\label{sec:study_data}
In the following evaluation experiments, we apply \softner{} to service incidents of \CompanyX{}, a major cloud service provider. These are incidents retrieved from large scale online service systems, which have been used by a wide distribution of users. In particular, we collected incidents spanning over a time period of 2 months. Each incident is described by its unique Id, title, description, last-modified date, owning team name and, also, whether the incident was resolved or not. Incident description is the unstructured text with an average of 472 words, showing us how verbose the incident descriptions are. Owning Team Name here, refers to the team to which the incident has been assigned.

\subsection{Entity Type Evaluation} 

Here, we evaluate the effectiveness of \softner{}'s unsupervised approach for named-entity extraction. Specifically, we evaluate the correctness of the entity types extracted by \softner{} on the entire study data. As the component performs \textbf{unsupervised} information extraction, we manually evaluate the precision of extraction. The component, first, extracts 100 distinct entities  sorted by the frequency of occurrence. We then, manually validate each potential entity and perform analysis on precision. Precision, here, is the fraction of extracted entities that are actually software entities.

\begin{figure}
\includegraphics[width=\linewidth]{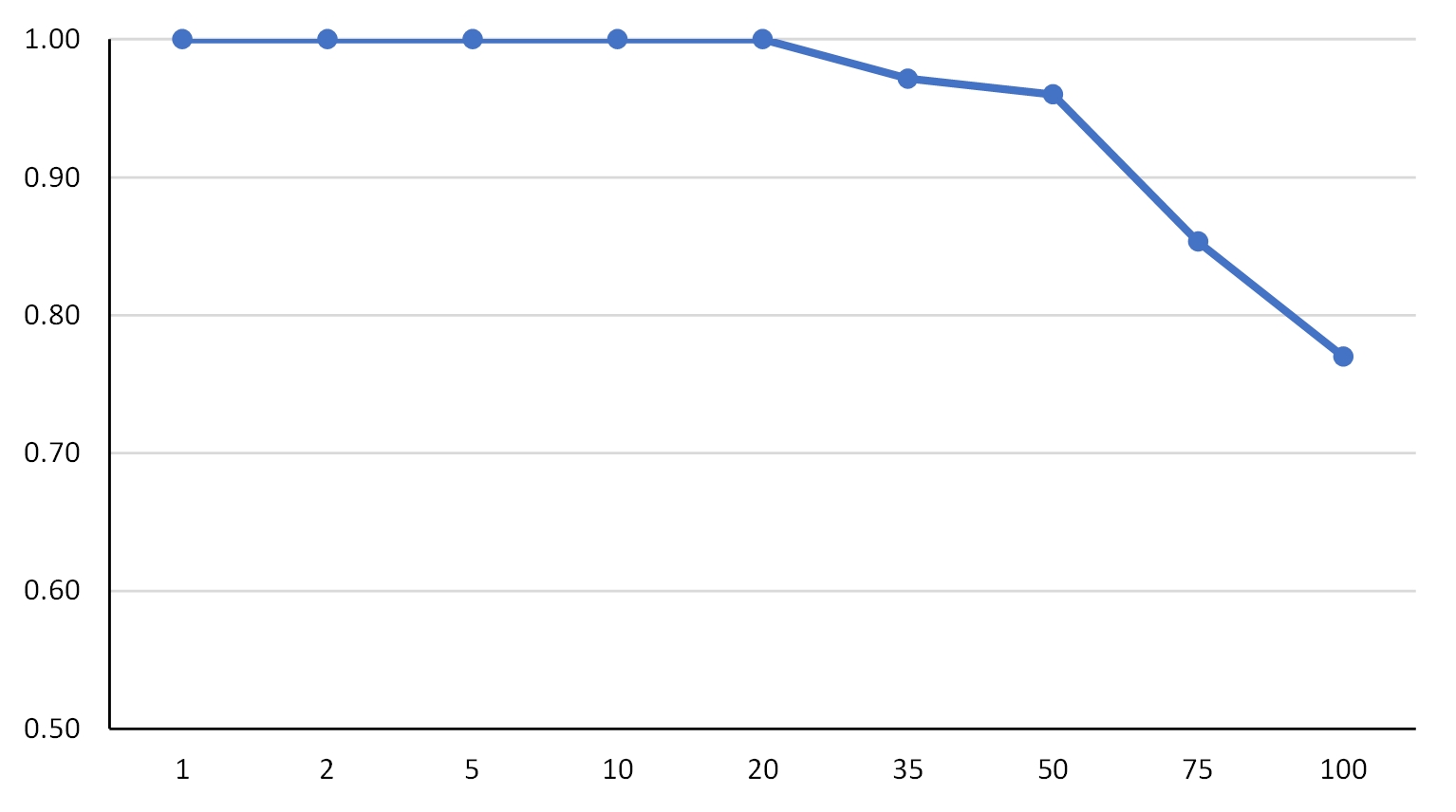}
\caption{Precision vs Rank curve for the entity types}
\label{pattern-extraction}
\end{figure}

Since the precision of \softner{}'s entity type extraction depends on the frequency of occurrence of entities, we further plot precision against a cut off rank $n$. Figure \ref{pattern-extraction} summarizes the precision of \softner{}'s entity type extraction against the top $n$ entities extracted, where $n \in [1,100]$. From this analysis, we see that \softner{} is able to extract 77 valid entities per 100 entities. We also see an expected decrease in precision, as $n$ increases, due to noisy tokens (false positives) like \textit{"to troubleshoot issue"}, \textit{"for cleanup delay"} and \textit{"for application gateway"}.

As explained above, in performing this experiment, n corresponds to the rank of the entity extracted. That is, a higher $n$ refers to an entity with low frequency of occurrence, which in turn can be extrapolated as an entity that is less important. \softner{}'s unsupervised entity type extraction has a minimal precision variation, also known as fall out rate, of 0.23 for an $n$ value as high as 100. This strengthens the hypothesis that \softner{}'s pattern extractors can pick up entities from unstructured text effectively, in a completely unsupervised manner.  

\subsection{\softner{} Model Evaluation}
Here, we evaluate the SoftNER deep learning model on the Named-Entity Recognition Task. We compare the multi task model, described in section \ref{sec:multi-task-model} and Figure \ref{model-arch}, against two baseline models, BiLSTM-CRF and BiLSTM-CRF with attention mechanism. These baseline models are state-of-the-art for NER \cite{huang2015bidirectional,lample2016neural, chiu2016named} and other NLP tasks as well. The models are compared on a fixed test set with 8300 incidents, that accounts for $20\%$ of the data-set. We use average precision, recall and F1 metrics to evaluate and compare the models on the NER task. The metrics are averaged over around 70 distinct type of entities tagged by the model. We also use a version of the F1 score, weighted based on the support for individual entities.

As shown in Table \ref{model-evaluation}, we observe that the baseline BiLSTM-CRF, with and without attention mechanism, achieves an average F1 score of around 0.88. Whereas, \softner{}'s Multi Task Model, as described in Section \ref{sec:multi-task-model}, achieves a higher average F1 score of around 0.96, i.e., a $\Delta F1\%$ of 8.7\%. \softner{} deployed as a solution in the incident management environment, is required to extract as much information from incident descriptions as possible. This is because, more information directly correlates with the ease of understanding the problem and identifying resources affected by the incident. We evaluate this ability using recall metrics and observe a high average recall of 0.95, as shown in Table \ref{model-evaluation}.

We further analyze the generalization of the model by analyzing test examples that were either false positive or false negative. Table \ref{false-examples} shows a few examples of sentences and the entities extracted from them. Note that we refer to false positives as FP, and, false negatives as FN in the table. We observe that some of the FPs are actually correct and were mislabelled in the test set because of the limitations of the pattern extractors. Let's take Example 1 from Table \ref{false-examples} for instance. Here, the unsupervised labelling component was only able to label \textbf{\textit{"2aa3abc0-7986-1abc-a98b-443fd7245e6"}} as \textbf{Subscription Id}, but not \textbf{\textit{"vaopn-uk-vnet-sc"}} as \textbf{Vnet Name} in the test sentence, due to restrictions with pattern extractors and label propagation. But the \softner{} model was able to extract both the entities from the sentence, proving it's ability to generalize beyond obvious structural pattern rules visible in the training data. In another example, row 2 in Table \ref{false-examples}, \textbf{403} was tagged \textbf{Error Returned} and \textbf{200} as \textbf{Status Code}, even though no structural elements (like ":") were present in the description. Row 3 shows a similar false positive example with the extraction of \textbf{\textit{192.168.0.5}} as \textbf{IP Address}. We also show a few contrasting false negatives, in rows 4 and 5, where the model was unable to extract entities \textbf{Ask} and \textbf{Ip Address} respectively.

\begin{table}
\small
\caption{Model evaluation}\vspace{-6pt}
\label{model-evaluation}
 \begin{tabular}[t]{lccc} 
\toprule
 \textbf{Metric} & \textbf{BiLSTM-CRF} & \textbf{BiLSTM-CRF} & \textbf{SoftNER} \\
  &  & \textbf{ Attention} & \textbf{Model} \\
 \midrule
 Avg F1 & 0.8803 & 0.8822 & \textbf{0.9572} \\
 Weighted Avg F1 &  0.9401 & 0.9440 & \textbf{0.9682} \\
 Avg Precision & 0.9160 & 0.9088 & \textbf{0.9693} \\
 Avg Recall &  0.8669 & 0.8764 & \textbf{0.9525} \\
 \bottomrule
\end{tabular}
\end{table}

\begin{table}
\small
\caption{FP and FN examples}\vspace{-6pt}
\label{false-examples}
\begin{tabular}[t]{ p{4cm} c p{2cm}} 
\toprule
\textbf{Sentence} & \textbf{Evaluation} & \textbf{Entities Tagged}\\
 & \textbf{Result} & \\
\midrule
\textit{SubscriptionId : 2aa3abc0-7986-1abc-a98b-443fd7245e6 unable to delete vnet name vaopn-uk-vnet-sc} & FP & {2aa3abc0-7986-1abc-a98b-443fd7245e6, vaopn-uk-vnet-sc }\\

\textit{Customer is reporting that their alerts for div-da are firing as 403 errors when the service is fully up and running with 200 codes} & FP & {403, 200} \\

\textit{Device Name : njb02-23gmk-isc, pa: 192.168.0.5 could not be configured! Error!} & FP & {njb02-23gmk-isc, 192.168.0.5} \\

\textit{The customer's  main ask: Need help to access cloud storage from cloud hosted service}  & FN & - \\

\textit{The loopback (ipv4) address (primary) is 192.131.75.235} & FN & {ipv4} \\
\bottomrule
\end{tabular}
\end{table}

\subsection{Auto-Triaging of incidents}
Incident triaging refers to the process of assigning a new incident to the responsible team. This is currently manually performed by the on-call engineers and it is not uncommon for the incident to be reassigned to a different team at a later point thereby reducing the accuracy and efficiency of incident management. These incidents often lead to huge economic losses and dissatisfaction amongst customers. Based on an empirical study, Chen et al. \cite{EmpiricalIcMICSE2019} showed that the reassignment rate for incidents can be as high as 91.58\% for online services at Microsoft. This shows the importance of incidents being assigned to the correct team at creation time. Several efforts \cite{EmpiricalIcMICSE2019, ContinuousTriageASE2019} have been made to automate the triaging process by leveraging the title, description and other meta-data of the incidents. Here, we evaluate the effectiveness of the knowledge extracted by \softner{} for the downstream task of automated triaging of incidents. Incident triaging is essentially a multi-class classification problem since the incident could be assigned to one of many teams. 

We sample 20\% of resolved incidents for the $10$ most common teams from the initial incident set (refer Section \ref{sec:study_data}) and run the \softner{} model on the description to extract the entities. The \softner{} entities can be broadly classified as either categorical or descriptive. While the descriptive entities are transformed to word embeddings using the same process described in Section \ref{sec:glove-word-emb}, the categorical entities are encoded into one-hot vectors. We then look at different combinations of features and compare the 5 fold cross-validation accuracy on various classification models. It is evident from Table \ref{tab:auto-triaging-models} that the models using the \softner{} entities, either on their own or along with the title, outperform the baseline models that use only the title and description information. We observe significant margins, with up to $7$\% - $27$\% increase in the 5 fold cross validation accuracy scores. These results reinforce that the entities extracted by \softner{} are indeed useful and can significantly help in downstream tasks. Using the entities extracted by \softner{} also reduces the input feature space since we no longer have to use the whole incident description. We also achieve high performance models using simple machine learning models thereby eliminating the need for complex deep learning models which have proven to be superior in past studies \cite{EmpiricalIcMICSE2019}.

In addition to comparing the cross validation accuracy of different models, we analysed feature significance by using the feature\_importances\_ attribute of a random forest model trained on the various input features. We observed that the entities extracted by \softner{} were given far more importance compared to the `Title', with the top features being - `exception message',  `problem type', `ask' and `issue' (as shown in Table \ref{tab:feature_significance}). This re-emphasises that the entities extracted from \softner{} boost the performance of classification models for the downstream task of automatic triaging of incidents.

\begin{table}
\small
    \begin{center}
    \caption{Comparison of 5 fold cross validation accuracy for auto-triaging using different feature sets.}
    \label{tab:auto-triaging-models}
    \vspace{-6pt}
    \setlength\tabcolsep{2 pt}
    \begin{tabular}{@{}p{2.66cm}ccccc@{}}
        \toprule
         
        \textbf{Feature Set}  & \textbf{Random} & \textbf{Linear} & \textbf{Gaussian} &  \textbf{K-Nearest} & \textbf{Naive} \\[-2pt]
         & \textbf{Forest} & \textbf{SVM} & \textbf{SVM} & \textbf{Neighbors} & \textbf{Bayes}\\
        \midrule
        Title + Description & 74.64 & 85.93 & 87.06 & 81.32 & 69.69 \\
        \midrule
        \softner{} Entities & 93.38 & 93.34 & 93.39 & 92.40 & 87.67\\
        $\Delta$ \% & 22.31 & 8.26 & 7.02 & 12.76 & 22.85 \\
        \midrule
        \softner{} Entities + Title & \textbf{98.60} & \textbf{99.20} & \textbf{98.95} & \textbf{99.14} & \textbf{88.07} \\
        $\Delta$ \% & 27.66 & 14.34 & 12.78 & 19.75 & 23.30 \\
        
        \bottomrule
    \end{tabular}
    \end{center}
\vspace{-6pt}
\end{table}

\begin{table}[H]
\vspace{-6pt}
\small\centering
    \caption{Importance scores for top features}
    \label{tab:feature_significance}
    \vspace{-6pt}
    \setlength\tabcolsep{2 pt}
    \begin{tabular}{lccccc}
        \toprule
        \textbf{Feature}  & Exception & Problem & Ask &  Issue & Title \\[-3pt]
        & Message & Type & & \\
        \midrule
        \textbf{Importance} & 0.0133 & 0.0111 & 0.0097 & 0.0051 & 0.0009\\
        \bottomrule
    \end{tabular}
\vspace{-6pt}
\end{table}
\section{Applications}
Automated extraction of structured knowledge from the service incidents unlocks several applications and scenarios:

\textbf{Incident summarization} - As shown in Table \ref{entity-examples}, \softner{} is able to extract key information from the incidents. This information can be added to the incidents as a summary for a quick overview for the on-call engineers. Otherwise, parsing the verbose incident descriptions to understand the issue and to locate the key information could be overwhelming. We have already enabled this scenario for one cloud service at \CompanyX{} and the feedback has been positive. We have added the summary for over 400 incidents.

\textbf{Automated health-checks} - Using \softner{}, we are able to extract key information such as \textit{Resource Id}, \textit{IP Addresses}, \textit{Subscription Id}, etc. This enables us to automate the manual tasks done by the on-call engineers for investigating the incidents. For instance, they will often pull up telemetry and logs for the affected resources. Or, they might look up the resource allocation for a given subscription. With the named entities, we can run these health checks automatically before the on-call engineers are engaged.

\textbf{Bug reports} - Even though this work is motivated by the various problems associated with incident management, the challenges with lack of structure applies to bug reports as well. We plan to evaluate \softner{} on bug reports at \CompanyX{} and, also, on the publicly available bug report datasets.

\textbf{Knowledge graph} - Currently, given an incident, \softner{} extracts a flattened list of entity name and value pairs. As a next step, we plan to work on relation extraction to better understand which entities are related. For instance, it will be quite useful to be able to map a given Identifier and IP Address to the same Virtual Machine resource. This would also allow us to incorporate the entities into a knowledge graph. This can unlock various applications such as entity disambiguation and linking.

\textbf{Better tooling} - The knowledge extracted by \softner{} can also be used to improve the existing incident reporting and management tools. We are able to build a named-entity set at any granularity (service, organization, feature team) in a fully unsupervised manner. These entities can then be incorporated into the incident report form as well, where some of these entities can even be made mandatory. We have already started working on this scenario with the feature teams which own incident reporting tools at \CompanyX{}.

\textbf{Type-aware models} - The mutli-task deep learning model architecture used by \softner{} uses both the semantic and the data-type context for entity extraction. As per our knowledge, this is the first usage of a multi-task and type-aware architecture in the software engineering domain. Given that software code and programs are typed, this model can potentially be used in other applications like code summarization where we can have a generative task along with a similar classification task for data-type prediction of the code tokens.

\textbf{Predictive tasks} - In this work, we have shown that the knowledge extracted by \softner{} can be used to build more accurate machine learning models for incident triaging. Similarly, we can build models to automate other tasks such as severity prediction, root causing, abstractive summarization \cite{paulus2017deep}, etc.
\section{Related Work}
Incident management has recently become an active area of research in the software engineering community. Significant work has been done on specific aspects of incident management, such as automated triaging of incidents, incident diagnosis and detection. Our work is complementary to the existing work on incident management since we focus on the fundamental problem of structured knowledge extraction from incidents. We also show that the knowledge extracted by \softner{} can be used to build more accurate models for these incident management tasks. \softner{} is inspired by the work on knowledge extraction in the Information Retrieval community. There have been decades of research on knowledge extraction from web data. However, majority of the research in the web space has been focused on supervised or semi-supervised entity extraction, limited to a known set of entities such as people, organizations and places. Also, unlike web pages, in software artifacts like incidents, the vocabulary is not limited to English and other human languages. Incidents contain not just textual information but also other entities such as GUIDs, Exceptions, IP Addresses, etc. Hence, \softner{} leverages a novel data-type aware deep learning model for knowledge extraction. Next, we discuss prior work in different domains which is related to this work.

\textbf{Incident management}: Recent work on incident management has been focused on the problem of triaging incidents to the correct teams. As per the empirical study by Chen et al. \cite{EmpiricalIcMICSE2019}, miss-triaging of incidents happen quite frequently and can lead up to a delay of over 10X in triaging time, apart from the lost revenue and customer impact. To solve this problem, they have also proposed DeepCT \cite{ContinuousTriageASE2019}, a deep learning based method for routing of incidents to the correct teams. They evaluate the model on several large-scale services and are able to achieve a high mean accuracy of up to 0.7. There has also been significant amount of work done on diagnosing and root causing of service incidents. Nair et al. \cite{nair2015learning} uses hierarchical detectors based on time-series anomaly detection for diagnosing incidents in services. DeCaf \cite{bansal2019decaf} uses random forest models for automatically detecting performance issues in large-scale cloud services. It also categorizes the detected issues into the buckets of new, known, regressed, resolved issues based on trend analysis. Systems such as AirAlert \cite{chen2019outage} have been built for predicting critical service incidents, called outages, in large scale services. It correlates signals and dependencies from across the cloud system and uses machine learning for predicting outages. Different from existing work, we focus on the fundamental problem of knowledge extraction from the service incidents. As we show in Section 5.4, using the named-entities extracted by \softner{}, we can build significantly more accurate models for these incident management tasks.

\textbf{Bug reports}: Significant amount of research has been done on bug reports in the traditional software context. \softner{} is inspired by InfoZilla \cite{bettenburg2008extracting} which leverages heuristics and regular expressions for extracting four elements from Eclipse bug reports: patches, stack traces, code and lists. Unlike InfoZilla, we build a completely unsupervised deep learning based framework for extracting entities from incidents. This enables \softner{} to extract hundreds of entities without requiring any prior knowledge about them. Our work also targets incidents, which are more complex than bugs because of numerous layers of dependencies and, also, the real-time mitigation requirements. Bug triage has been an active area of research \cite{anvik2006should, tian2016learning, bortis2013porchlight, wang2014fixercache} in the software engineering community. Other aspects of bug reports such as bug classification \cite{zhou2016combining} and bug fix prediction time \cite{ardimento2017knowledge} have also been explored. Similar to incidents, existing work on bug reports have largely used the unstructured attributes like bug description as it is. Even though in this work, we have focused on incidents, \softner{} can be applied to bug reports for extracting structured information and building models for tasks like triaging, classification, etc.

\textbf{Information retrieval}: Knowledge and entity extraction has been studied in depth by the information retrieval community. Search engines like Google and Bing rely heavily on entity knowledge bases for tasks like intent understanding \cite{pantel2012mining} and query reformulation \cite{xu2008entity}. Several supervised and semi-supervised methods have been proposed for entity extraction from the web corpora \cite{florian2003named, mccallum2003early, pacsca2007weakly, vyas2009semi}. Supervised methods require large amount of training data which can be cost prohibitive to collect. Hence, the search engines commonly use semi-supervised methods which leverage a small seed set to bootstrap the entity extraction process. For instance, the expert editors would seed the entity list for a particular entity type, let's say fruits with some initial values such as \{apple, mango, orange\}. Then using pattern or distributional extractors, the list would be expanded to cover the entire list of fruits. In this work, our goal was to build a fully unsupervised system where we don't need any pre-existing list of entity types or seed values. This is primarily because every service and organization is unique and manually bootstrapping \softner{} would be very laborious. Also, unlike web pages, incident reports are not composed of well-formed and grammatically correct sentences. Hence, we propose a novel multi-task based deep learning model which uses both the semantic context and the data-type of the tokens for entity extraction.
\section{Conclusion}
Incident management is a key part of building and operating large-scale cloud services. Prior work on incident management has focused on specific aspects such as triaging, predicting incidents and root causing. In this paper, we propose \softner{}, a deep learning based unsupervised framework for knowledge extraction from incidents. \softner{} uses structural patterns and label propagation for bootstrapping the training data. It then incorporates a novel multi-task BiLSTM-CRF model for automated extraction of named-entities from the incident descriptions. We have evaluated \softner{} on the incident data from \CompanyX{}, a major cloud service provider. Our evaluation shows that even though \softner{} is fully unsupervised, it has a high precision of 0.96 (at rank 50) for learning entity types from the unstructured incident data. Further, our multi-task model architecture outperforms existing state of the art models in entity extraction. It is able to achieve an average F1 score of 0.957 and a weighted average F1 score of 0.968. As shown by the manual evaluation, \softner{} is also able to generalize beyond the structural patterns which were used to bootstrap. We have deployed \softner{} at \CompanyX{}, where it has been used for knowledge extraction from over 400 incidents. We also discuss several real-world applications of knowledge extraction. Lastly, we show that the extracted knowledge can be used for building significantly more accurate models for critical incident management tasks like triaging.

%
\bibliographystyle{ACM-Reference-Format}
\bibliography{references}

\end{document}